\documentclass[12pt]{article}
\usepackage{graphicx}
\usepackage{scicite}

\usepackage{times}

\newenvironment{sciabstract}{%
\begin{quote} \bf}
{\end{quote}}

\newcounter{lastnote}

\title{Probing Asymmetric Molecules with High Harmonic Generation. [Original manuscript as prepared on 22/05/2011]}

\author
{E. Frumker,$^{1,3,4}$ N. Kajumba,$^{1,4,5}$ J. B. Bertrand,$^{1}$ H. J. W\"{o}rner,$^{1,6}$\\
 C. T. Hebeisen,$^{1}$ P. Hockett,$^{2}$ M. Spanner,$^{2}$ S. Patchkovskii,$^{2}$\\
  G. G. Paulus,$^3$ D. M. Villeneuve,$^1$ P. B. Corkum$^1$\\
\\
\normalsize{$^{1}$Joint Attosecond Science Laboratory, University of Ottawa and National Research,}\\
\normalsize{Council of Canada, 100 Sussex Drive, Ottawa, On, Canada}\\
\normalsize{$^{2}$Steacie Institute for Molecular Science, National Research Council of Canada}\\
\normalsize{100 Sussex Drive, Ottawa, On, Canada}\\
\normalsize{$^{3}$ Department of Physics, Texas A\&M University, College Station,}\\
\normalsize{Texas 77843, USA}\\
\normalsize{$^{4}$ Max-Planck-Institut f\"{u}r Quantenoptik, Hans-Kopfermann-Strasse 1,}\\
\normalsize{ D-85748 Garching, Germany}\\
\normalsize{$^{5}$Department f\"{u}r Physik der Ludwig-Maximilians-Universit\"{a}t, Schellingstrasse 4,}\\
\normalsize{D-80799 Munich, Germany}\\
\normalsize{$^{6}$ Laboratorium f\"{u}r physikalische Chemie, ETH Z\"{u}rich,Wolfgang-Pauli-Strasse 10,}\\
\normalsize{ 8093 Z\"{u}rich, Switzerland}\\
\\
}

\date{}

\begin{document}


\baselineskip24pt


\maketitle


\begin{sciabstract}
  Asymmetric molecules look different when viewed from one side or the other. This difference influences the electronic structure of the valence electrons, thereby giving stereo sensitivity to chemistry and biology.  We show that attosecond and re-collision science provides a detailed and sensitive probe of electronic asymmetry. On each $1/2$ cycle of an intense light pulse, laser-induced tunnelling extracts an electron wave packet from the molecule. When the electron wave packet recombines, alternately from one side of the molecule or the other, its amplitude and phase asymmetry determines the even and odd harmonics radiation that it generates. This harmonic spectrum encodes three manifestations of asymmetry; an amplitude and phase asymmetry in electron tunneling; an asymmetry in the phase that the electron wave packet accumulates relative to the ion between the moment of ionization and recombination; and an asymmetry in the amplitude and phase of the transition moment. We report the first measurement of high harmonics from oriented gas samples.  We determine the phase asymmetry of the attosecond XUV pulses emitted when an electron recollides from opposite sides of the CO molecule, and the phase asymmetry of the recollision electron just before recombination. We discuss how the various contributions to asymmetry can be isolated in future experiments.
\end{sciabstract}


High harmonic and attosecond science gives us direct access to the electronic time scale and electronic phase.  In single photon ionization in atoms, for example, photoelectrons from different bound states are found to ionize with different time delays \cite{Schultze_Delay_Photoemission_Science2011}.  In other words, the global phase structure of the transition moment across the full bandwidth of the photoelectron depends on the initial state.  In another example, high-order multiphoton ionization, approximated by tunneling, imposes a different amplitude and phase on tunneling electrons depending on the bound states from which the electron is extracted \cite{Smirnova_multielectron_Nature2009}.  In both of these cases, one state serves as a reference for the other.  In fact, this "internal interferometer" is exploited when high harmonic spectroscopy is used to time resolve photochemical dynamics \cite{Woerner_Chem_Nature_2011}.  Unexcited molecules serve as a local oscillator against which dissociation is coherently resolved.

 The process of high harmonic generation can be understood as cascading interferometers.  This is the basis of the exquisite sensitivity to phase asymmetry and makes high harmonics especially suited for asymmetric molecules. Here we use a single state of an asymmetric molecule and one side as a reference for the other. In the first interferometer, removal of the electron by the electric field of an intense light pulse quantum mechanically splits the electron wavefunction into two parts; one part is pulled away from the molecule by the oscillating laser field and then is driven back where it interferes with the asymmetric bound part of the same wavefunction.  The interference produces an attosecond XUV pulse.  In the second interferometer, the temporal periodicity of the laser field creates an attosecond pulse train that leads to discrete frequency components in the emission spectrum.  For a symmetric molecule, the alternating parity of each attosecond burst results in odd harmonics of the laser frequency.  Phase matching -- the coherent addition of the signals from many molecules -- is like a third interferometer.   For an asymmetric medium both even and odd harmonics will be measured. In other words, measuring the harmonics signal reads the second interferometer which in turn reads the first as illustrated in Fig. 1. Many aspects of molecular asymmetry are thereby opened to measurement.

While molecular frame measurements have demonstrated that the multiphoton ionization probability depends on orientation \cite{Akagi_tunnel_ion_Science2009}, we create an oriented gas in laboratory frame of sufficient density for harmonic emission and measure the characteristic even harmonic signal expected for a system without inversion symmetry \cite{Boyd_NLO}. (Following common usage, we refer to orientation when a specific atom of an asymmetric molecule preferentially points in a specific laboratory direction. We refer to the case where parallel and anti-parallel are equivalent as alignment). We calculate the amplitude asymmetry of tunneling and recombination. This allows us to determine the frequency-dependent total phase asymmetry - an important and highly sensitive new characteristic of an asymmetric molecule. We then calculate the phase asymmetry of the transition moment using scattering electron wavefunctions (please see SOM (IIA)) \cite{Le_QRS_PRA2009}. Combining experimental and theoretical results we determine the phase asymmetry of the recollision wave packet at the moment of recollision.

We work with the isoelectronic pair of $\mathrm{N_2}$ and $\mathrm{CO}$ (ionization potential $\mathrm (IP) = 15.6$eV and $14.0$eV respectively, and the dipole moment $(\mathrm{D_{CO}}) = 0.112$Debye) \cite{TRICKL_N2_IP_JCP1989,ERMAN_CO_IP_CPL1993,Burrus_D_CO_JCP1958}. The electronic structure of their Dyson highest occupied molecular orbitals (HOMO) is shown in Fig. 2(a). We choose this pair because experiments with aligned molecules have shown that the net ionization probability when averaged over orientations as a function of angle is similar \cite{Pinkham_Ionization_CO_PRA2005,Alnaser_Symmetries_Dissociation_PRA2005}.

Either gas was injected into a vacuum system using a supersonic jet as described in SOM (I(a)). We achieve alignment and orientation using a pulse of $800$nm light and its second harmonic \cite{De_Orientation_w2w_PRL2009}. Together they serve as a pump pulse, aligning and orienting the sample. After a variable delay, we probe the sample with an intense 800 nm pulse to create high harmonics. By delaying the probe, we observe the sample at different stages of alignment and orientation.  The optical setup is described in SOM (I(a)).

We begin with the time dependence of the alignment and orientation wave packet.
Figure 3 (upper panel) shows the time evolution of a typical odd harmonic signal (H19; $\mathrm{h\nu \sim29eV}$) for $\mathrm{N_2}$ (dashed blue line). The origin of the time axis for $\mathrm{N_2}$ (and CO) is taken as the time of the first full rotational revival for each molecule.  For $\mathrm{N_2}$, the odd harmonic signal peaks when the molecule is aligned \cite{Itatani_HHG_wavepackets_PRL2005,Velotta_HHG_aligned_2001}.  This only changes at extreme intensities for the cut-off harmonics due to the contribution of the HOMO-1 orbital \cite{McFarland_N2_HOMO_1_Science2008}.  Figure 3 (upper panel) also shows the 19th harmonic for $\mathrm{CO}$ (solid red line). Comparing $\mathrm{N_2}$ and $\mathrm{CO}$, we see a different alignment structure - for $\mathrm{CO}$ the odd harmonic signal peaks when the molecule is perpendicular. This is true for all harmonics that we have studied except for the very lowest (H11 and H13).  For comparison, we also show in the dotted green curve in Fig. 3 (upper panel) the alignment of $\mathrm{CO}$ as measured by Coulomb explosion imaging (please see SOM (I(b) for details). $\mathrm{CO}$ is optimally aligned at about the time of the H19 minimum.

There are two reasons why the harmonic signal can peak when a molecule is perpendicular to the driving laser field. First, the ionization probability depends on the ionization potential \cite{Keldysh_ionization_JETP1965}. Levels with relatively similar ionization potential can ionize and contribute to the harmonic emission \cite{McFarland_N2_HOMO_1_Science2008}.  In addition, the ionization probability is modulated by the momentum wavefunction \cite{Muth_Bohm_Faisal__molecular_ion_PRL2000,Perelomov_PPT2_JETP1967,Tong_LinCD_Molecular_ADK_PRA2002} of the state from which the electron tunnels.   In both $\mathrm{CO}$ and $\mathrm{N_2}$, the node in the  $\pi_u$ HOMO-1 minimizes this orbital's contribution when the molecule is parallel.   However, the HOMO and HOMO-1 contribute almost equally to the high harmonic emission (please see SOM Fig. 2) when $\mathrm{CO}$ is perpendicular, enhancing the high harmonic yield for anti-aligned molecules.

The second reason is unique to asymmetric molecules. When the field is parallel to the CO axis, successive electron wave packets can be quite different. This asymmetry leads to even harmonics and imperfect constructive (or even destructive) spectral interference for odd harmonics. Contrast this with the attosecond XUV pulse train created from perpendicular molecules.  Now all pulses in the train interfere perfectly to maximize odd harmonics.  The reduction of the odd harmonic signal along the direction of asymmetry persists even when the sample as a whole is symmetric. We will see that this contribution is important in CO. We now turn to macroscopically oriented samples, where even harmonics become observable.

Figure 3 (lower panel) shows the time-dependent orientation of $\mathrm{CO}$ as determined by Coulomb explosion imaging (dashed green curve).  Orientation is measured by the cosine of the angle between an axis parallel to the electric field of the two-colour laser pulse and the momentum (or inverse momentum) of the $\mathrm{C}^{++}$ (or $\mathrm{O}^{++}$) fragments (please see SOM I(b)). The time dependence of a typical even harmonic (H14; $\mathrm{h\nu \sim 22eV}$; solid red line) maximizes approximately when the orientation also maximizes. The even harmonics appear on zero background allowing a signal-to-noise better than $500:1$.  (Please note:  The CEI measurement was made using a different jet and in a different chamber.  Therefore, small differences in revival dynamics are expected.  What is important is that both the even harmonic signal and orientation maximize together.)

Figure 4 shows the full harmonics spectrum taken at a pump-probe time delay of $8.85$ps - the time of maximum orientation.  A spectrum of even harmonics is clearly visible as are the much stronger odd harmonics. This spectrum measures molecular asymmetry as seen by the recollision electron.  Therefore, it is important, although challenging, to quantify the ratio of odd to even harmonics over the large dynamic range in the figure.  All data presented in Fig. 4 have been calibrated following the procedure presented in SOM (I(A)).

In Figure 4, the upper curve is a line-out of the data in the experimental image just below.  The ratio of intensities of each even harmonic order to the average of the adjacent odd harmonic orders is plotted as the upper of the two bar graphs.  The ratio increases up to a photon energy of $28$eV (H18), where the trend reverses. It has a clear minimum around a photon energy of $37$eV (H24) and then grows monotonically to the cut-off.

Having introduced the idea of imperfect constructive interference, it is also important to present the alignment behavior of all odd harmonics.  The lower bar graph shows the normalized difference, $R=(H_{aa}-H_{a})/H_{b}$ of the odd harmonics where the subscripts represent aligned (a), anti-aligned (aa) and baseline (b). We measure $H_b$ after the revival has dephased.  We will draw conclusions from this figure, but first we turn to theory.
Each step of the three step process of high harmonic generation \cite{Corkum_3step_PRL1993} (the first interferometer) offers a unique perspective on molecular asymmetry, some of which we can estimate.

\underline{Tunneling:}  During its birth, tunneling introduces an asymmetry in the amplitude ratio, $r_t=A^O_t/A^C_t$ and a phase difference, $\delta\phi_t=\phi^O_t- \phi^C_t$ for electron wave packets created from opposite sides of the molecule (here $A^O_t$, $\phi^O_t$, $A^C_t$ and  $\phi^C_t$ are amplitudes and phases of tunneling from the $-\mathrm{O}-$ and $-\mathrm{C}-$ sides of $\mathrm{CO}$, respectively). Asymmetry arises for at least three reasons.  First, as shown in Fig. 2(a), the ionizing orbital is asymmetric.  This leads to an amplitude asymmetry \cite{Akagi_tunnel_ion_Science2009}.  Second, the Stark shift of the neutral molecule and the molecular ion makes the ionization potential direction dependent.  This also leads to an amplitude asymmetry \cite{Akagi_tunnel_ion_Science2009,Etches_Madsen_Stark_Orientation_JPhysB2010}.  Third, the departing wave packet interacts with other electrons (also asymmetrically arranged) through Coulomb and exchange forces, creating a phase asymmetry.  Estimating the tunneling phase (or its asymmetry) remains an unsolved problem \cite{Smirnova_multielectron_Nature2009}. However, we can estimate the amplitude asymmetry of the ionization probability, $r_t$ (shown in Fig. 2(b)) by following the computational procedure described in SOM (IIb).  Like for $\mathrm{N_2}$ \cite{Tong_LinCD_Molecular_ADK_PRA2002,Litvinyuk_N2_ion_PRL_2003} the probability of ionization for $\mathrm{CO}$ maximizes when the molecule is aligned \cite{Pinkham_Ionization_CO_PRA2005,Alnaser_Symmetries_Dissociation_PRA2005}, but it is highly asymmetric.  The ionization rate peaks when the electric field points towards the $\mathrm{-C-}$ atom, similar to the behavior of $-\mathrm{HCl}-$ towards the $-\mathrm{H}-$ atom \cite{Akagi_tunnel_ion_Science2009} and $-\mathrm{OCS}-$ towards the $-\mathrm{S}-$ atom \cite{Stapelfeldt_OCS_orient_NatPhys2010}.  	

\underline{Wave packet propagation:} Once the wave packet is launched by tunneling, phase accumulates rapidly relative to the Stark-shifted ion ground state.  Asymmetry primarily arises because of the permanent dipole. While propagation will have little impact on the re-collision amplitude asymmetry (i.e. $r_p \approx 1$), phase asymmetry, $\delta\phi_p$, will depend on the harmonic order. The $\delta\phi_p$ can be estimated within the strong field approximation and it is shown in SOM Fig. 5.

\underline{Recombination:}  The final step describes the recollision electron making a transition to its initial bound state.  Quantum chemistry methods allow us to calculate the field-free transition moment, shown in Fig. 2(c) for $\mathrm{CO}$.  Both amplitude (plotted in the third dimension) and phase (represented by color) depend on the photoelectron's direction relative to the molecule leading to an amplitude ratio $r_r$ and a phase difference $\delta\phi_p$. The calculation of the transition moment is described in SOM (IIA).
Additional asymmetry of the transition moment will arise from the asymmetric polarization of the electronic density caused by the laser field. Estimation of this contribution has never been addressed.

Taken together, the product of the amplitude ratio of the tunneling wave packet (Fig. 2(a)) and the amplitude ratio of the transition moment (Fig. 2(c)) determines the total amplitude ratio of attosecond bursts emitted from the $-\mathrm{C}-$ vs. $-\mathrm{O}-$ side.
This information, together with an estimate of the degree of orientation, determines the phase asymmetry of successive attosecond XUV pulses from the measured even/odd harmonic amplitude ratio in Figure 4.
 The reconstructed phase asymmetry ( as described in SOM (IID)) is plotted in the inset in Figure 4 as the red curve.
This phase asymmetry of the XUV bursts includes the phase asymmetry of the transition moment and of the recollision electron wave packet. It is a characteristic of the molecule and the field. Under the same laser conditions, the phase asymmetry will be different for different molecules.

We can remove the contribution of the phase of the transition moment, $\delta\phi_r$, shown as the dotted curve in Fig. 4 inset, thereby isolating the phase asymmetry of the attosecond electron wave packet at the moment of recollision ($\delta\phi_\omega=\delta\phi_{CO}-\delta\phi_r$). An important feature of the reconstructed phase is the zero crossing of all three curves near $37$eV (H24). The zero crossing correlates with the minimum in the even to odd harmonic ratio. In other words, near $37$eV the phase asymmetry that the electron gains by tunneling (a previously unmeasured parameter) is approximately equal to the phase asymmetry that the electron wave packet gains along its trajectory between ionization and re-collision.

With the relative phase of the adjacent attosecond XUV and electron bursts measured, we return to the normalized difference for the odd harmonics in Figure 4.  The measured phase asymmetry of successive XUV pulses is minimal for aligned molecules near $\mathrm{\sim37 eV}$ (H25). Near $37$eV successive attosecond bursts contribute optimally to the odd (H25) harmonics leading to a minimum in the modulation depth in Fig. 4 (bottom panel). The minimum in both the normalized difference and in the even/odd harmonic ratio share a common origin in the near total phase symmetry at $\mathrm{\sim37 eV}$.

Our calculations allow us to expand upon the issue of delayed single photon ionization \cite{Schultze_Delay_Photoemission_Science2011}.  The transition dipole in Figure 2(c) shows that photoionization occurs at different times for electrons departing in different directions from a single state of an asymmetric molecule. On the C-side, the spectral phase is approximately the same for all frequencies.  On the $\mathrm{O}$-side, there is a nonlinear phase sweep.  A photoelectron's wave packet is not only delayed, but also distorted on the $\mathrm{O}$-side (please see SOM Fig. 6).

Looking forward, it is feasible to systematically isolate each manifestation of molecular asymmetry. We have only measured the amplitude of the harmonics at one orientation.  Existing techniques allow molecular frame measurements \cite{Sakai_interf_CO2_Nature2005,Vozzi_2center_PRL2005} for all orientations as well as to measure the relative spectral phase of all harmonics over a wide frequency range (the equivalent to measuring a time delay) \cite{Paul_Rabbit_Science2001,Kanai_HHG_mixed_gases_PRL_2007}.

To progress further, we need to manipulate one or more of the cascading interferometers.  There are many approaches . For example, by independently controlling the wavelength and intensity of the fundamental laser beam, the propagation time of the electron in the continuum can be changed without changing the electric field at the moment of electron tunneling and without changing the energy of re-collision. This will isolate the propagation contribution (primarily due to Stark shift), while keeping tunneling and recombination constant. Similar manipulation will allow all parameters to be isolated.
We could also add a new interferometer to the cascade.  One way is to add a weak second harmonic component to the probe pulse \cite{Liu_Watanabe_2color_PRA_2006,Mauritsson_pulse_train_2color_PRL2006,Dudovich_w2w_situ_NaturePhys2006}. In such 2-color experiments one makes "even harmonics" (even with respect to the fundamental field) by adding or subtracting one or more second harmonic photons during the highly nonlinear harmonic generation process. Controlling the relative phase between $\omega+2\omega$ adds a controllable even harmonic source against which the phase of even harmonics from oriented molecules can be studied.

Already, high harmonic spectroscopy of asymmetric molecules presents a major challenge for the theory of photoionization and high harmonic generation. Never before have HHG experiments provided such an interferometrically sensitive measure of the transition moment phase or of the phase of the recollision electron wave packet.  Since the even harmonic spectrum encodes {\it differences} in the phases and amplitudes of the attosecond bursts from opposite sides of the molecule with interferometric sensitivity, unprecedented accuracy on the computation of each step in the HHG process will be necessary for capturing the correct asymmetries leading to the even HHG spectrum.  The strong field approximation - the most common approximation used in high harmonic generation - will not be up to this challenge.  In the paper we have stressed that tunneling models have barely addressed the issue of a tunneling phase and its associated asymmetry.  Further, models used to calculate transition moments may need improvement, with likely inclusion of subtle asymmetries arising from field-dependent asymmetric polarization.  What is clear is that the qualitative experimental differences in structure of the even harmonic spectrum relative to the odd harmonics will provide new insight into high harmonic generation and into the molecules that we use.
%

Figure 1.

Spectral interference lies at the heart of high harmonic generation from oriented molecules.  Every $1/2$ period of the fundamental intense infrared laser pulse (shown in red), an electron wave packet is detached from an oriented molecule, oscillates in the time-dependent electric field and recombines to create an attosecond burst of XUV light.  An example of four such bursts of a train is shown in the Figure. In our experiment, this attosecond pulse train is diffracted and imaged by the grating (see SOM (IA) for the details). The spectral interference of such a train results in the formation of even and odd harmonics. An experimental spectrum is shown in the Figure.

Figure 2.

(a) The Dyson orbitals for the most weakly bound valence electron of the isoelectronic pair, $\mathrm{N_2}$ and $\mathrm{CO}$. The colour code represents a $\pi$ phase difference. The orientation is preserved in (b) and (c).  (b) The total ionization yield as a function of the angle between the molecular axis and the electrical field calculated for a $1/2$ cycle of $800nm$ with intensity  $1.5*10^{14}W/cm^2$.  (c) The amplitude (vertical axis) and phase (color code) of the transition moment as a function of angle and harmonic order (radial axis) for XUV light polarized along the molecular axis.

Figure 3.

Upper plot: A rotational wave packet in CO excited by impulsive alignment with an 800 nm pulse is imaged by Coulomb explosion in the dotted green curve.  Using this rotational wave packet the dashed blue curve shows the time dependence of the high harmonic spectrum for the 19th harmonic for $\mathrm{N_2}$,  and $\mathrm{CO}$ (solid red line) near the time of their first full revival (set at t=0). Lower plot: An orientational wave packet in CO excited by the superposition of 800 nm and 400 nm light is measured by Coulomb explosion in the dotted green curve.  The red curve shows the time dependent signal for the 14th harmonic.  Note that the even harmonic signal is maximum when the orientation is greatest.

Figure 4.

The high harmonic spectrum is shown at $8.85$fs, the time of maximum orientation. The intensity of all harmonics is presented in the top panel of the figure, showing in graphical form what is measured in the image below.  The ratio of even to the average of the adjacent odd harmonics is shown in the upper bar graph. The lower bar graph plots the normalized difference -  $R=(H_{aa}-H_{a})/H_{b})$.  The figure inset shows the phase difference (in green, circle marks) for an electron departing from the $\mathrm{-C-}$ side and the $\mathrm{-O-}$ side of the molecule as determined by e-polyscat simulations. The red (cross marks) curve is the measured spectral phase asymmetry of successive attosecond XUV (recollision electron) pulses (relative to the normally expected 180 degrees asymmetry). The blue (square marks) curve is the recolliding electron wave-packet spectral phase asymmetry.


%
\bibliographystyle{unsrt}

\clearpage
 \begin{figure}[t]
	\centering
	\includegraphics[width=\columnwidth]{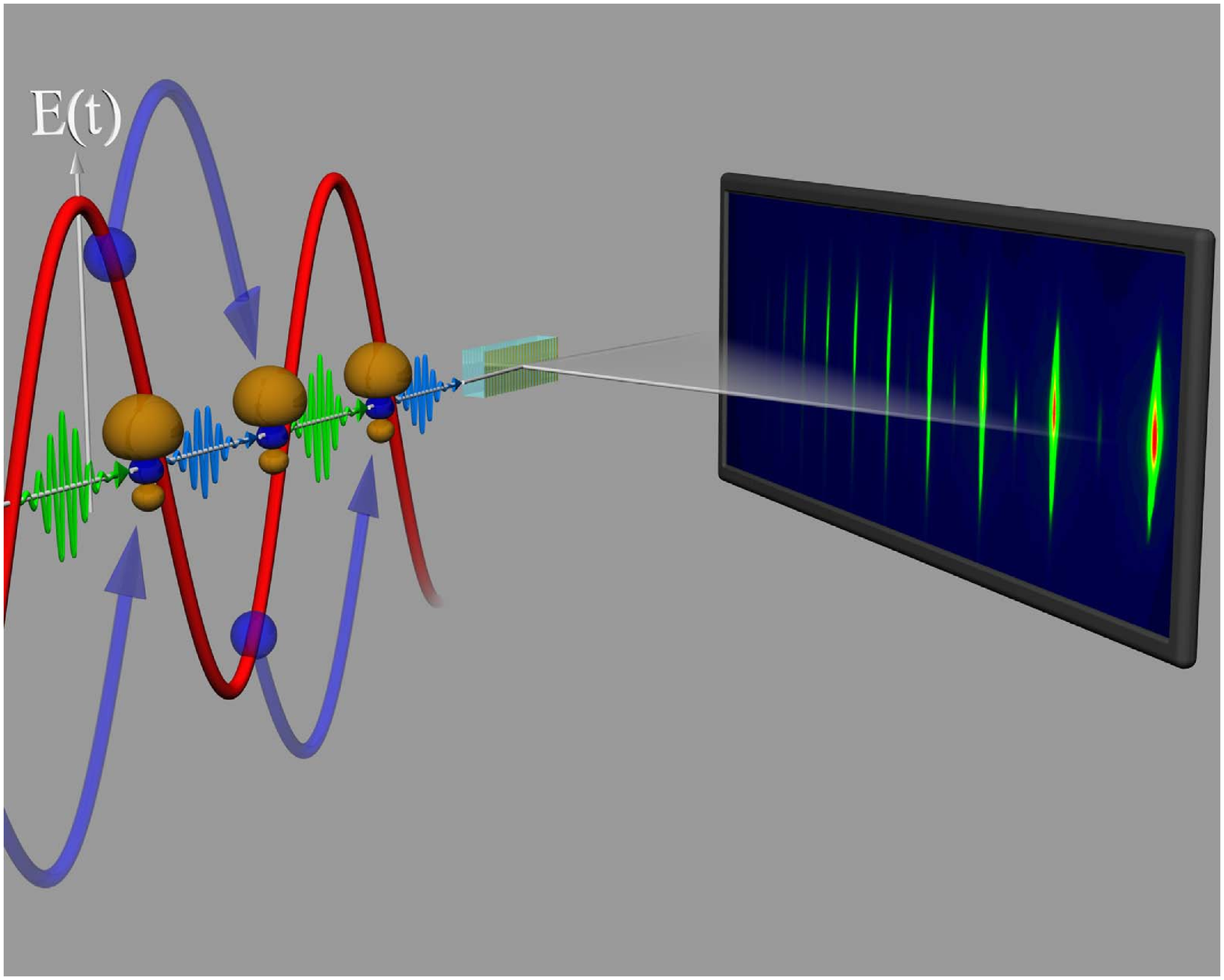}
	\caption{Spectral interference lies at the heart of high harmonic generation from oriented molecules.  Every $1/2$ period of the fundamental intense infrared laser pulse (shown in red), an electron wave packet is detached from an oriented molecule, oscillates in the time-dependent electric field and recombines to create an attosecond burst of XUV light.  An example of four such bursts of a train is shown in the Figure. In our experiment, this attosecond pulse train is diffracted and imaged by the grating (see SOM (IA) for the details). The spectral interference of such a train results in the formation of even and odd harmonics. An experimental spectrum is shown in the Figure.}
	\label{Fig1}
\end{figure}

\begin{figure}[t]
	\centering
	\includegraphics[width=\columnwidth]{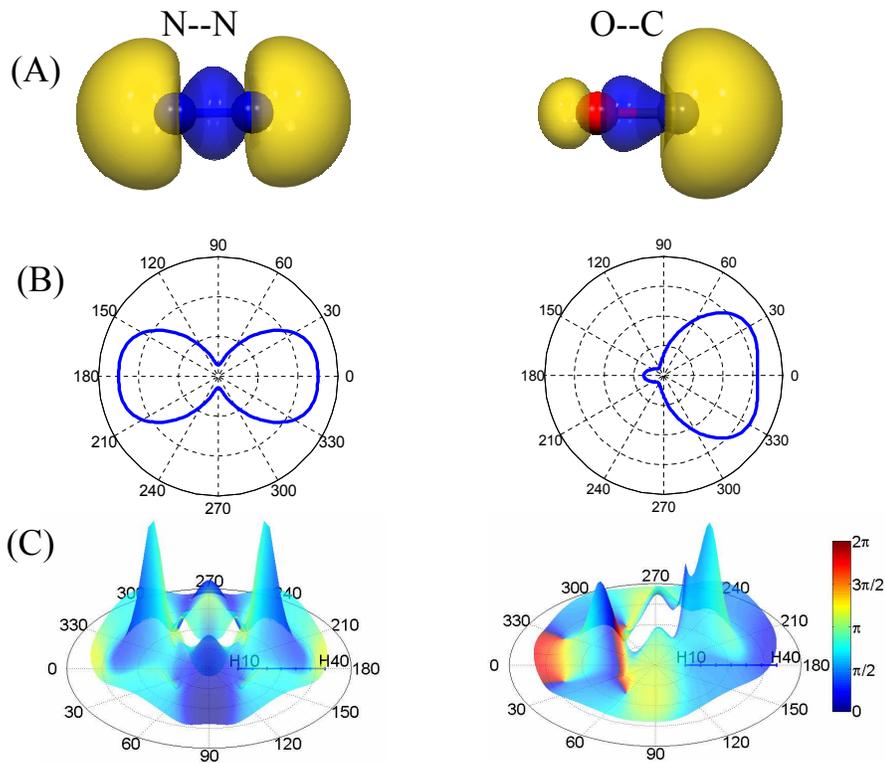}
	\caption{(a) The Dyson orbital for the most weakly bound valence electron of the isoelectronic pair, $\mathrm{N_2}$ and $\mathrm{CO}$. The colour code represents a $\pi$ phase difference. The orientation is preserved in (b) and (c).  (b) The total ionization yield as a function of the angle between the molecular axis and the electrical field calculated for a $1/2$ cycle of $800nm$ with intensity  $1.5*10^{14}W/cm^2$.  (c) The amplitude (vertical axis) and phase (color code) of the transition moment as a function of angle and harmonic order (radial axis) for XUV light polarized along the molecular axis.}
	\label{Fig2}
\end{figure}

\begin{figure}[t]
	\centering
	\includegraphics[width=\columnwidth]{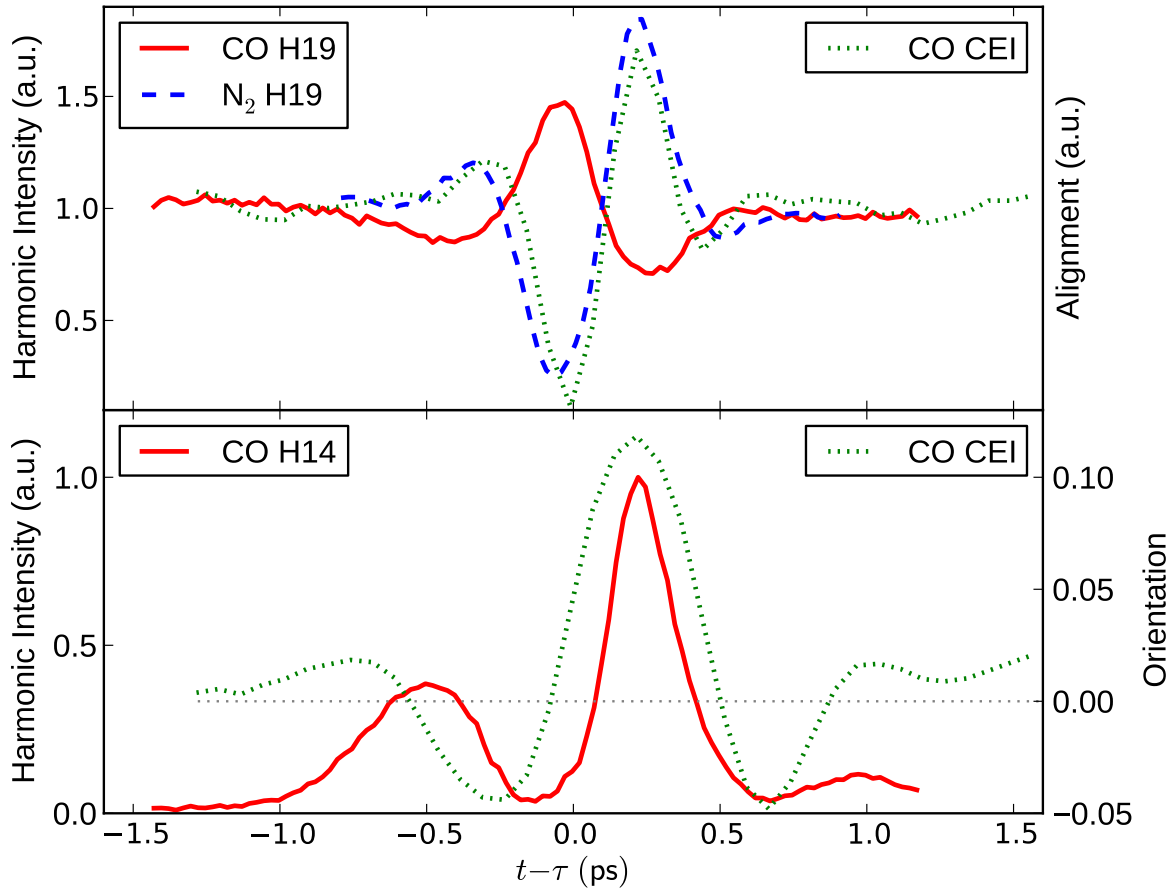}
	\caption{Upper plot: A rotational wave packet in CO excited by impulsive alignment with an 800 nm pulse is imaged by Coulomb explosion in the dotted green curve.  Using this rotational wave packet the dashed blue curve shows the time dependence of the high harmonic spectrum for the 19th harmonic for $\mathrm{N_2}$,  and $\mathrm{CO}$ (solid red line) near the time of their first full revival (set at t=0). Lower plot: An orientational wave packet in CO excited by the superposition of 800 nm and 400 nm light is measured by Coulomb explosion in the dotted green curve.  The red curve shows the time dependent signal for the 14th harmonic.  Note that the even harmonic signal is maximum when the orientation is greatest.}
	\label{Fig3}
\end{figure}

\begin{figure}[t]
	\centering
	\includegraphics[width=\columnwidth]{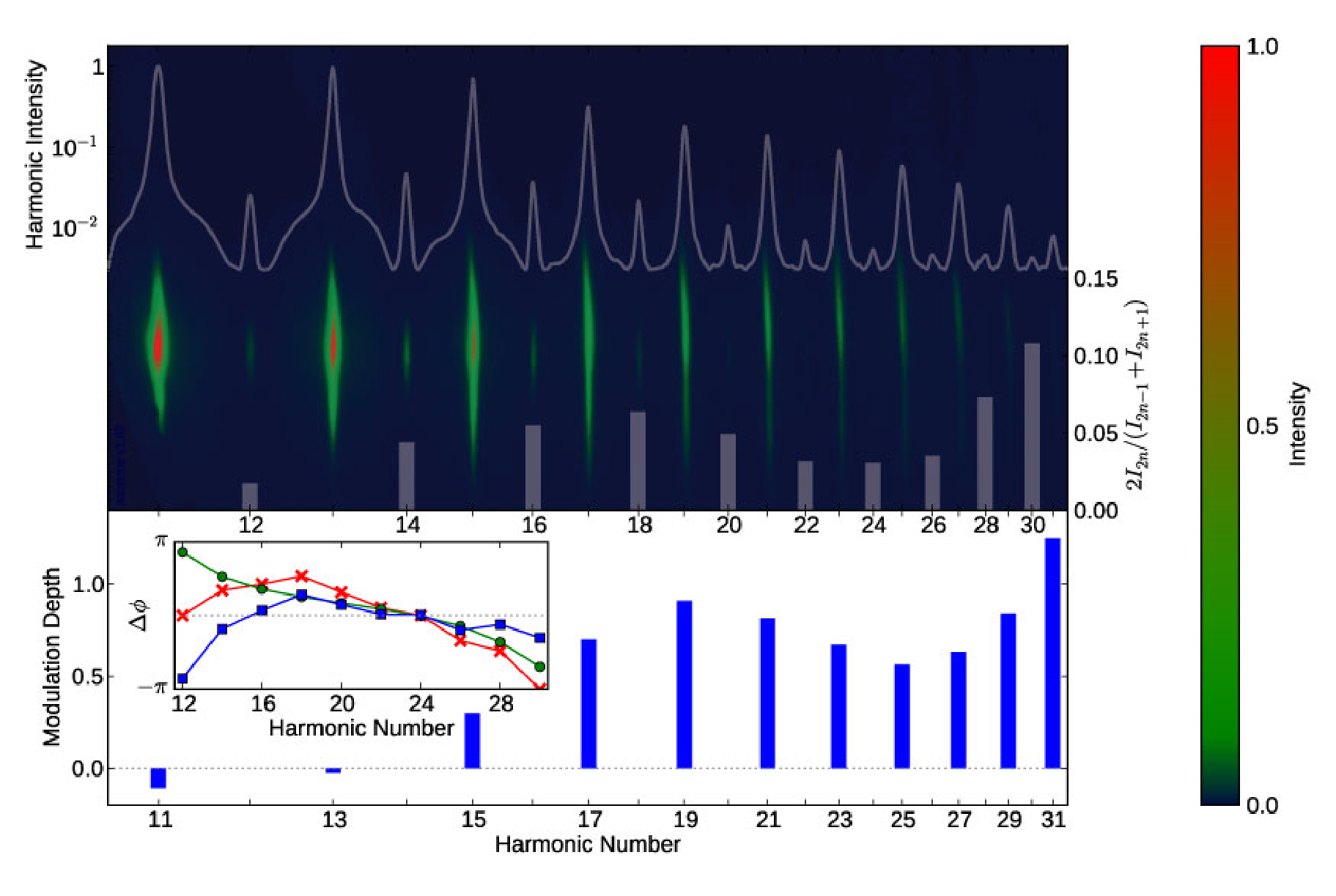}
	\caption{
The high harmonic spectrum is shown at $8.85$fs, the time of maximum orientation. The intensity of all harmonics is presented in the top panel of the figure, showing in graphical form what is measured in the image below.  The ratio of even to the average of the adjacent odd harmonics is shown in the upper bar graph. The lower bar graph plots the normalized difference -  $R=(H_{aa}-H_{a})/H_{b})$.  The figure inset shows the phase difference (in green, circle marks) for an electron departing from the $\mathrm{-C-}$ side and the $\mathrm{-O-}$ side of the molecule as determined by e-polyscat simulations. The red (cross marks) curve is the measured spectral phase asymmetry of successive attosecond XUV (recollision electron) pulses (relative to the normally expected 180 degrees asymmetry). The blue (square marks) curve is the recolliding electron wave-packet spectral phase asymmetry.}
	\label{Fig4}
\end{figure}

\end{document}


\title{Supporting Online Materials. [Original version as prepared on 22/05/2011]}

\maketitle

\section{Experimental Details}

\subsection{High harmonics measurements}

A linearly polarized laser pulse from a Titanium:Sapphire laser ($\sim$35 fs,
$\lambda$=800 nm, 50 Hz, $\sim$10 mJ) was split into pump and probe arms. A
$\lambda$/2 waveplate and a thin film polarizer were used in each arm to
continuously control the pulse energies.

The pump beam serves to produce impulsive alignment and orientation.
 We introduce asymmetry in the electric field of the
pump pulse, by mixing the fundamental ($\omega$) beam with its second harmonic
(2$\omega$) and controlling the phase relationship between the two frequencies.
The second harmonic beam ($\lambda$=400 nm) was generated in a Type-I BBO crystal
(thickness 200 $\mu$m), so that its polarization
was perpendicular to the fundamental beam. The pump beam was de-magnified (1.5:1 telescope) to ensure full overlap with the probe beam when focused on the jet.  The conversion efficiency was $\sim$10\%.
The fundamental and second harmonic beams co-propagate through a birefringent
calcite plate to precompensate for the group velocity mismatch between the
fundamental (800 nm) and the second harmonic (400 nm) pulses as they propagate
through the optical elements and the air downstream. A zero order $\lambda$/2 waveplate
at 45$^\circ$ was used to bring the beams' polarization into the same (vertical)
plane. Phase control between the $\omega$ and 2$\omega$ beams
was achieved by adjusting the angle of a 1 mm fused silica window placed in the beam
after the waveplate.

The high-harmonic generating (probe) beam was delayed using a
computer-controlled delay line.
The parallel, but spatially separated, pump and probe beams impinge on a spherical f=500 mm
focusing mirror.  The beams overlap on the gas jet with a small angle (≈1$^\circ$)
between their propagation directions. The probe beam was focused before the gas
jet. This configuration favors short trajectory harmonics over long
trajectory harmonics. A pulsed valve (Parker pulse valve, $\sim$200 $\mu$s pulse
duration, 40 psi back pressure, 100 $\mu$m aperture) was used to produce the
supersonic gas jet synchronized with the laser pulses. The gas was 10\% CO,
diluted in He, for enhanced cooling. The non-collinear beam geometry ensures
spatial separation of the harmonics produced by the alignment/orientation
(pump) beam and by the harmonic generating (probe) beam, respectively.  Therefore, the
even harmonics produced through the field asymmetry in the pump pulse
do not contribute any background signal to the weak even harmonics produced by
the probe pulse in the partially oriented gas sample. The generated harmonics were
analyzed in an XUV flat field spectrometer. The details of this spectrometer
are described elsewhere \cite{Frumker_SWORD_OL2011}. The high harmonics spectrum was
detected using an imaging multi-channel plate (MCP) / phosphor screen detector.
Images from the phosphor screen were recorded by a cooled charge-coupled device
(CCD) camera.

To ensure the greatest possible linearity of the detector and to
facilitate extended dynamic range data acquisition, the MCP detector was
operated at a low voltage (1700 V across the MCP chevron pair) and the same
image was acquired by the cooled CCD multiple times with different integration
times. We determined the linear range of the camera with respect to the
integrated intensity and masked out saturated and non-linear regions of each
image.  The remaining parts of the images were integrated and calibrated
according to the total acquisition time of each pixel.  This method provides
substantial noise reduction and an extended dynamic range over single
acquisitions.

\subsection{Coulomb Explosion Imaging measurements}

For Coulomb explosion measurements we used a different chamber and a different laser but the same method to achieve orientation.
The pump beam size was reduced by a 2:1 telescope and a 300
$\mu$m BBO crystal generates the second harmonic light (22\% conversion
efficiency).  A calcite window again compensates for the group delay between
the two wavelengths and a $\lambda$/2 (at 800 nm) waveplate brings the
polarization of the fundamental pulse into the same (horizontal) plane as that
of the second harmonic.  A rotatable 1 mm thick fused silica window was used to
control the relative phase.  The probe pulse was a horizontally polarized 800 nm
pulse.  Its delay relative to the pump pulse was set by a computer-controlled
delay stage.  The two parallel beams were focused by a spherical mirror (f=50
mm) on a continuous gas jet of pure CO in the centre of the Coulomb explosion
Imaging (CEI) spectrometer, which is described elsewhere \cite{Dooley}.

The spectrometer detects the molecular fragments ionized by the laser
pulses.  From their position on the detector and the
time-of-flight information gathered by the detector, the 3D momentum of a
fragment can be reconstructed along with its charge-to-mass ratio.  In a
diatomic molecule, the momentum of a fragment resulting from Coulomb explosion
is in the direction of the molecular axis.  We used the momenta of C++ and O++
ions to reconstruct the orientations of the fragmented molecules.  The linearly
oriented probe pulse preferentially ionizes molecules whose molecular axis is
near-parallel with the laser polarization.  Hence, the detected molecular
orientation shows a bias towards alignment along the horizontal axis i.e.
$\langle \cos^2\vartheta\rangle$ is much higher than expected for an evenly
probed sample.  However, since the probe beam's electric field does not exhibit
asymmetry, there is no preference in the orientation signal and $\langle
\cos\vartheta\rangle$ reflects the net preferential orientation of molecules
whose axis is near-parallel to the probe polarization.

\section{Theoretical analysis and numerical calculations}

\subsection{Calculation of recombination dipole matrix elements}
\label{Section_recomb}
The field-free recombination dipole matrix elements shown in Fig. 1 of the main
text were calculated using ePolyScat \cite{Lucchese1,Lucchese2}, followed by
further post-processing.  In general the dipole matrix elements for molecular
targets, formulated as a scattering matrix element involving a bound state and
a continuum function, are complex quantities which are non-trivial to evaluate.
The difficulty arises from the calculation of the continuum wavefunction, which
must extend over all space, fulfill the necessary asymptotic boundary
conditions and take full account of the multi-polar nature of the molecular
potential.  ePolyScat uses Hartree Fock input from standard electronic
structure codes to define the bound states, and then uses the Schwinger
variational method to calculate the continuum electron state.  We use the
GAMESS \cite{GAMESS} package with the (built-in) aug-cc-pVDZ basis set
\cite{Dunning}.  Dipole matrix elements are calculated and output in terms of
an expansion in spherical waves.

The dipole matrix elements for photoionization can be expressed as \cite{Lucchese2,CDLin}
\begin{equation}
	d_{{\mathbf k},\hat n} = k^{1/2}\langle \Psi_i|{\mathbf r}\cdot\hat n|\Psi^{(-)}_{f,{\mathbf k}} \rangle
\end{equation}
where $\Psi^{(-)}_{f,{\mathbf k}}$ is the final state of the system, including
the bound ion states and continuum electron; ${\mathbf r}\cdot\hat n$ is the
dipole operator, with $\hat n$ defining the dipole radiation polarization and
${\mathbf r}$ the electronic coordinates; $\Psi_i$ is the initial bound state.
This can be rewritten in terms of a spherical wave expansion of the continuum
\cite{Lucchese2,CDLin}
\begin{equation}
	d_{{\mathbf k},\hat n} = \left[ \frac{4\pi}{3} \right]^{1/2}
	\sum_{l,m,\nu} d_{l,m,\nu} Y_{l,m}(\hat k) Y_{1,\nu}(\hat n)
\end{equation}
The outputs from ePolyScat are the $d_{l,m,\nu}$, the radial part of the dipole
matrix elements indexed by partial wave at a given energy.  Here $l$ and $m$
index the partial waves, $\nu$ indexes the components of the dipole radiation
expressed in a spherical basis, $\hat k$ is the unit wavevector for the free
electron and $\hat n$ the unit polarization vector of the dipole radiation.
The photoionization matrix elements described above are the complex-conjugate
of the recombination matrix elements required in high harmonic calculations \cite{CDLin}.

Experimentally the emitted high harmonic yield is measured as a
function of $\theta$, the angle between the driving laser field and the
molecular axis.  The calculated (molecular frame) recombination dipoles can be
transformed into the frame-of-reference of the driving field (the lab frame) by
a simple frame rotation.  Furthermore, we assume that only the parallel
component of harmonics is required, corresponding to the case where the continuum
electron is ejected, driven, and recombines along the polarization direction of
the driving field, i.e. $\hat k \parallel \hat n' \parallel \hat n$ where $\hat
n'$ is the driving field polarization vector.  Under this simplification the
emitted radiation at an angle $\theta$ is given by
\begin{equation}
	d_{\hat k \parallel \hat n' \parallel \hat n}(\theta) = \cos(\theta)d^{\parallel}_{{\mathbf k},\hat n}
	+ \sin(\theta)d^{\perp}_{{\mathbf k},\hat n}
\end{equation}
where $d^{\parallel}_{{\mathbf k},\hat n}$  and $d^{\perp}_{{\mathbf k},\hat
n}$ are the required elements of the molecular frame dipole matrix elements for
$\hat k \parallel \hat n' \parallel \hat n$, and are referenced to the
molecular axis by their superscripts.

The recombination dipole maps in Fig.1 (main text) show many salient features of the
angle-dependence of high-harmonic generation. N$_2$ shows, as
expected for a homonuclear molecule, mirror symmetry in the matrix elements.
The matrix elements peak along the molecular axis, but also show significant
recombination amplitude perpendicular to the N-N bond.  As a function of energy
of the recombining electron (hence harmonic order) both the magnitudes and
phases of the matrix elements vary, although this effect is most pronounced
along the molecular axis.

The dipoles for CO is asymmetric, as expected for
a heteronuclear diatomic. Specifically, the recombination amplitude at the C
end is greater than the O end.  Importantly, significant phase differences in
the recombination dipoles are also shown in the calculation, with a difference
of around $\pi/2$ between the C and O recombination at the peak amplitudes (around
harmonic 20).  This phase difference in the recombination matrix elements plays an important role in the
phase difference of the emitted attosecond bursts from opposite ends of an
oriented system.

\subsection{Calculations of the angle-dependent ionization probability}
\label{Section_ionization}

We calculated angle-dependent ionization yields using the single-channel
version of the time-dependent numerical {\it ab initio} method outlined in
Ref.\cite{SpannerPatchkovskii}.  This method uses multielectron quantum
chemistry wavefunctions to represent the multielectron bound states of the
neutral and cation, and couples these states to Cartesian grids used to
represent the continuum states.  The method includes effects of the ionic core
on the outgoing electron, and does not assume tunneling or semiclassical
ionization formulas. For the present calculations we computed the ground state
of the neutral and ionic systems using GAMESS \cite{GAMESS} with the
correlation-consistent polarized valence triple zeta (cc-pVTZ) basis set
\cite{Dunning} at the complete active space (CAS) level using 10 (neutral) or 9
(cation) active electrons in 8 orbitals.  The continuum grid extended to
$\pm$13 a.u. in all directions with a grid step size of 0.2 a.u.  The
time-dependent equations of Ref.\cite{SpannerPatchkovskii} were integrated
using the leap-frog method with a time step of 0.0026(6) a.u.

The system is initiated with all population in the neutral ground state and then exposed to a
half-cycle of 800 nm light of intensity $1.5\times10^{14}$ W/cm$^2$. As the laser field interactions with our model system and neutral population is coupled to the ionic state, population builds
up on the continuum grid and flows away from the core. We use absorbing boundaries \cite{Manolopoulos} with a width of 4.3 a.u.  at
the edges of the continuum grid to prevent reflection of the outgoing electron
flux and to calculate the ionization yield by monitoring the population
absorbed at the grid edges.  The calculation was repeated with a range of
different angles between the molecular axis and the laser polarization
direction in order to capture the full angular dependence of the ionization
yield. The results appear in Figure 2 in the text.

\subsection{Calculation of numerical high harmonic spectra}

\begin{figure}[t]
	\centering
	\includegraphics[width=\columnwidth]{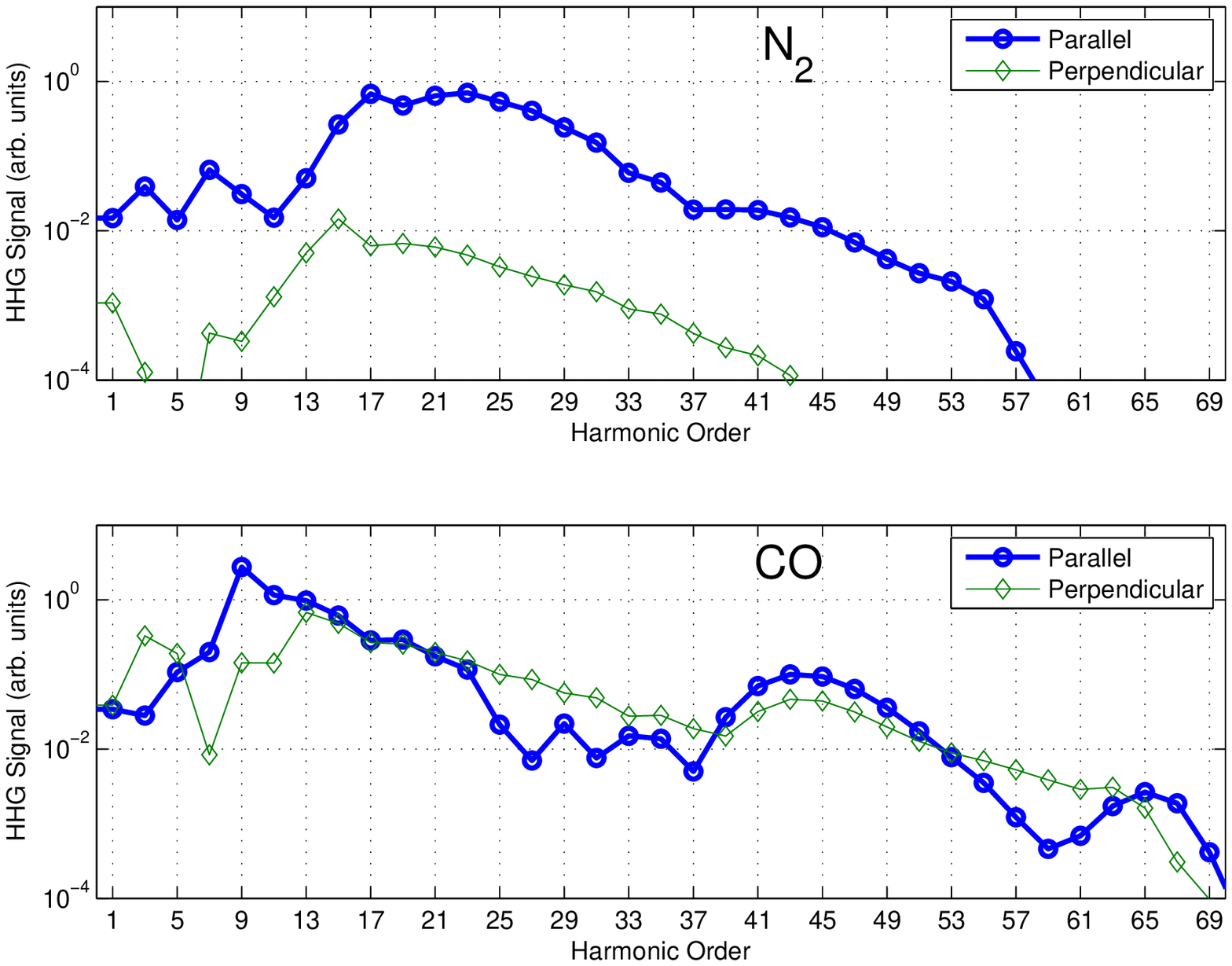}
	\caption{Numerical high harmonic spectra of the HOMO channel of  N$_2$ (top) and CO (bottom)
	at $2.8\times10^{14}$ W/cm$^2$.}
	\label{FigN2CO}
\end{figure}

\begin{figure}[t]
	\centering
	\includegraphics[width=\columnwidth]{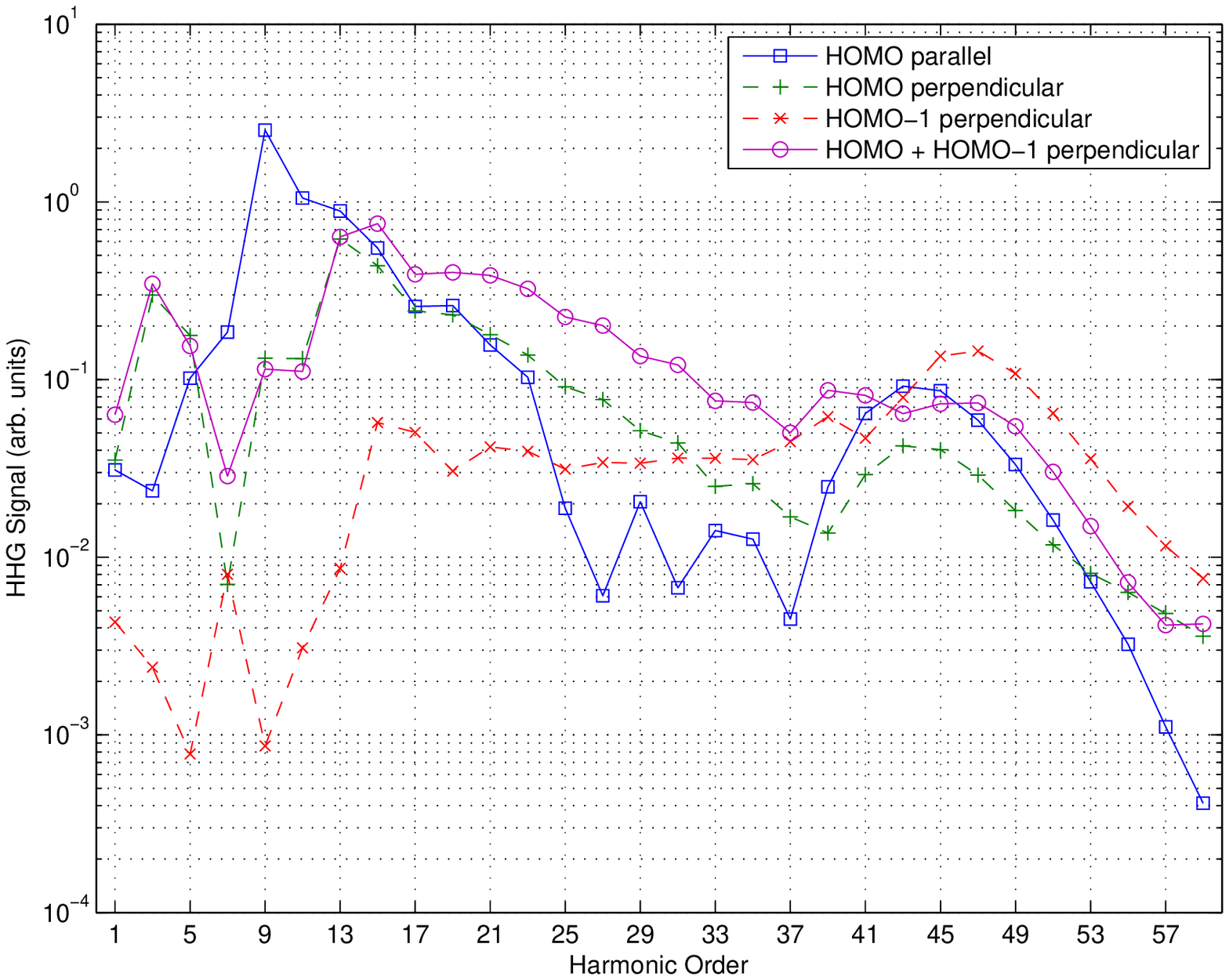}
	\caption{Numerical high harmonic spectra of CO demonstrating the contribution of HOMO-1 in the
	perpendicular configuration.}
	\label{FigCO-homo-homo-1}
\end{figure}

\begin{figure}[t]
	\centering
	\includegraphics[width=\columnwidth]{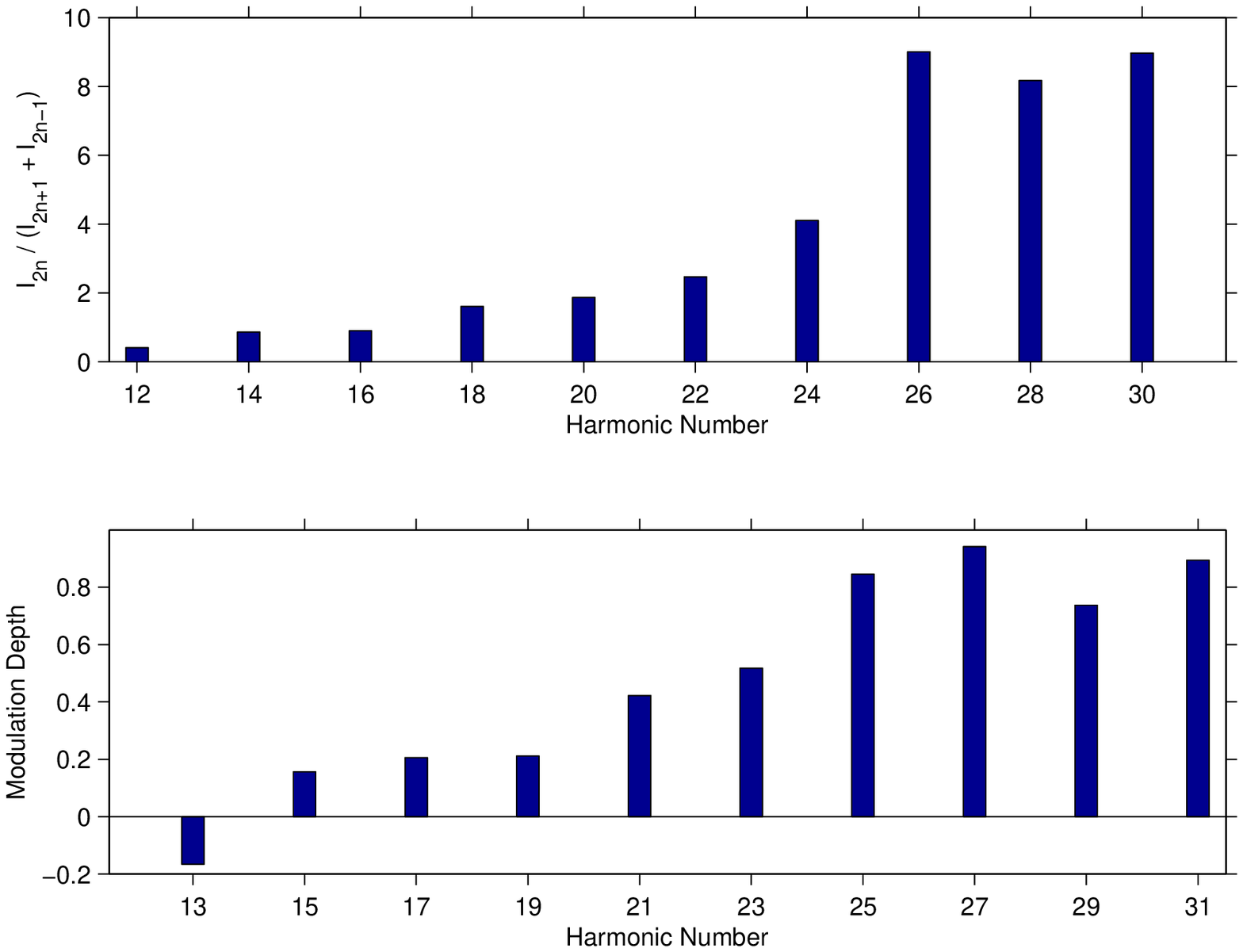}
	\caption{Ratio of even-to-odd harmonics and modulation depth for simulated CO harmonics.}
	\label{FigCObars}
\end{figure}

In addition to ionization yields, we can also calculate the high harmonic
spectrum using the computational method outlined in
Ref.\cite{SpannerPatchkovskii}.  The calculation requires a
grid large enough to allow the continuum wavepacket to propagate away from the
core some distance before returning to the core for recollision. Thus we
need a bigger grids to calculate harmonics than was needed for ionization.

We use a grid extending from $\pm$17.5 a.u. in directions perpendicular
to the laser polarization axis and $\pm$42.5 a.u. along the laser polarization
axis.  The space step size was 0.213 a.u. and the time step size was again
0.002666 a.u. The absorbing boundaries were used with a width of 8.2 a.u. in
all directions. In order to calculate the harmonic spectrum,
we use the Fourier transform of the dipole expectation value
\begin{equation}
	S_{HHG}(\omega) = \omega^4 \left|\int dt\: d(t) e^{i\omega t}\right|^2
\end{equation}
where
\begin{equation}
	d(t)= \langle \Psi(t)|\vec r\cdot \vec e_L|\Psi(t)\rangle
\end{equation}
and $\vec e_L$ is the laser polarization direction.
In the notation of Ref.\cite{SpannerPatchkovskii} the dipole becomes
\begin{equation}
	d(t)= \sqrt{n} \left[ a_m^*(t) \langle \tilde \phi
^S_m | \vec r_n | \chi_m(t)\rangle
	+{\cal N}_{\tilde N} b^*(t)\langle \vec\phi^C_m|\chi_m(t)\rangle \right]\cdot \vec e_L  + c.c.
\end{equation}
where $c.c.$ means complex conjugate (see Ref.\cite{SpannerPatchkovskii} for
specific definition of all symbols in this equation).  The dipole $d(t)$ was
calculated for a simulation running 2.5 cycles of monochromatic (no envelope
was used) 800 nm light.  The Fourier transform used to calculate
$S_{HHG}(\omega)$ included only the final 2 cycles of $d(t)$, thus minimizing
assymetries due to the initial turn on of the field.  In addition, we used a
Welch window when implementing the numerical Digital Fourier Transform to avoid discontinuities of
$d(t)$ at the edges of the time window.

Figure \ref{FigN2CO} shows the numerically simulated high harmonic spectra for the HOMO
channels of N$_2$ and CO at a peak intensity of $2.8\times10^{14}$ W/cm$^2$.
Two configurations are shown; the parallel case where the molecular axis is
along the polarization direction of the laser and the perpendicular case where
the molecular axis is orthogonal to the laser polarization.  (For CO, harmonics in
the parallel case are calculated by coherently adding the contributions from
'0$^\circ$' and '180$^\circ$' configurations.)  For N$_2$, the perpendicular
case is roughly 2 orders of magnitude below the parallel case indicating no
inversion of the harmonic alignment revivals.  However, for CO the harmonics in the
perpendicular geometry are not only on the same order as the parallel harmonics
across the spectrum, but they even dominate for a larger region indicating
inversion of the harmonic alignment revivals.  With the present grid sizes, the
finer features of the spectrum are not fully converged and show a small
dependence on the exact position of the nuclei relative to the computation
grid.  However, the general trend that the parallel harmonics in N$_2$ always
dominate the perpendicular harmonics while in CO the perpendicular harmonics
are seen to overwhelm the parallel harmonics is stable.

The HOMO-1 orbital of CO has a maximum that lies perpendicular to the molecular
axis and a node along the molecular axis.  For this reason it is possible that
HOMO-1 plays a role in the perpendicular configuration for CO.  Figure
\ref{FigCO-homo-homo-1} investigates the contribution of HOMO-1.  Shown are the
parallel (blue) and perpendicular (green) configurations of the HOMO channel,
the same harmonic spectra as the bottom panel of Fig.~\ref{FigN2CO}.  In addition we
now plot the harmonic signal from the HOMO-1 channel in the perpendicular
configuration (red) and the coherent sum of the HOMO and HOMO-1 channels in the
perpendicular configuration. (In the parallel configuration, the HOMO-1 channel
harmonics were multiple orders of magnitude weaker than the HOMO channel harmonics).
It is seen that the perpendicular HOMO-1 harmonics are on the same order as the perpendicular
HOMO channel.  The HOMO and HOMO-1 channels together now display a wide range of
harmonics over which inversion of the alignment revivals is seen.

The experimental harmonic data is presented as the ratio  of even-to-odd
harmonics and the modulation depth (Fig.4 of the main text).  Figure
\ref{FigCObars} shows the corresponding simulated values for these two
measures.  The even-to-odd ratios of harmonic intensities is extracted from the
simulated harmonic spectra by considering a fully oriented sample, the '0$^\circ$'
case alone.  The simulated modulation depth calculated by using the HOMO
channel harmonics in the parallel (aligned) configuration and the coherent sum
of the HOMO and HOMO-1 channels for the perpendicular (anti-aligned) case.
Although the simulations fail to reproduce the minima seen in the experimental
measures, the general trend for these measures to increase across the measured
harmonic range is captured.

\subsection{Phase reconstruction}

 Knowledge of amplitude asymmetry, together with estimation of the degree-of-orientation, allows us to determine the phase asymmetry of successive attosecond XUV pulses in the train from the measured even/odd harmonic amplitude ratio - $r(\nu_{2n})$ (shown in Fig. 4 in the body of the paper). To exemplify our approach, let's assume a simple model in which molecules are perfectly aligned and only partially oriented (with $n_\uparrow$ molecules pointing up and $n_\downarrow$ pointing down). In this case the degree-of-orientation is defined as $\alpha_o \equiv (n_\uparrow - n_\downarrow)/(n_\uparrow + n_\downarrow)$. For a single polar molecule, emitting attosecond bursts with the spectral amplitudes $E_C(\nu)=|E_C(\nu)|\exp(i\varphi_C(\nu))$ and $E_O(\nu)=|E_O(\nu)|\exp(i\varphi_O(\nu))$, the intensity of emitted even harmonics is proportional to  $|E_C(\nu)-E_O(\nu)|^2$, while intensity of the odd harmonics is proportional to $|E_C(\nu)+E_O(\nu)|^2$. Within this model the spectral phase asymmetry ($\triangle\varphi=\varphi_C-\varphi_O$) can be approximated as


\begin{equation}
\triangle\varphi(\nu_{2n})\simeq\pm\arccos\left[\frac{1+ \eta(\nu_{2n})}{ 2\sqrt\eta(\nu_{2n})} \cdot \left(1-\frac{2r(\nu_{2n})}{r(\nu_{2n})+\alpha_o^2}\right)\right] \label{Eq_phase_retrieval_single_molecule}
\end{equation}

where $\eta(\nu_{2n})$ - is the spectral intensity ratio of  $-\mathrm{C}-$ vs.  $-\mathrm{O}-$ attosecond bursts, which we calculate based on asymmetry in ionization rate (Sections \ref{Section_ionization}) and in amplitude of recombination dipole (Section \ref{Section_recomb}).

We experimentally estimate $\alpha_o$ in two ways. (1) In our Coulomb explosion experiments we measure the degree-of-orientation to be $\alpha_o~\sim 0.12$.
The CEI apparatus has a low backing pressure continuous gas jet. For the high harmonic experiment we use a pulsed jet with $\sim3\textrm{bar}$ backing pressure and He buffer gas. The even harmonic signal increases with backing pressure and buffer gas. Therefore $\alpha_o$ measured in CEI is probably an underestimate.\\ (2) The HHG is a coherent process, therefore the measured signal is proportional to the square of the number of emitters. This implies that the ratio of even to the average of the adjacent odd harmonics
$r(\nu_{2n})$ would be proportional to the square of the degree-of-orientation. Assuming that the spectral phase between attosecond bursts changes significantly across the spectrum, we may estimate the degree-of-orientation as an average value of $r(\nu_{2n})$, shown in the upper bar graph in Figure 4 (main text).
This average value is around $\sim0.05$, which implies a degree-of-orientation $\alpha_o\sim0.22$.

\begin{figure}[t]
	\centering
	\includegraphics[width=\columnwidth]{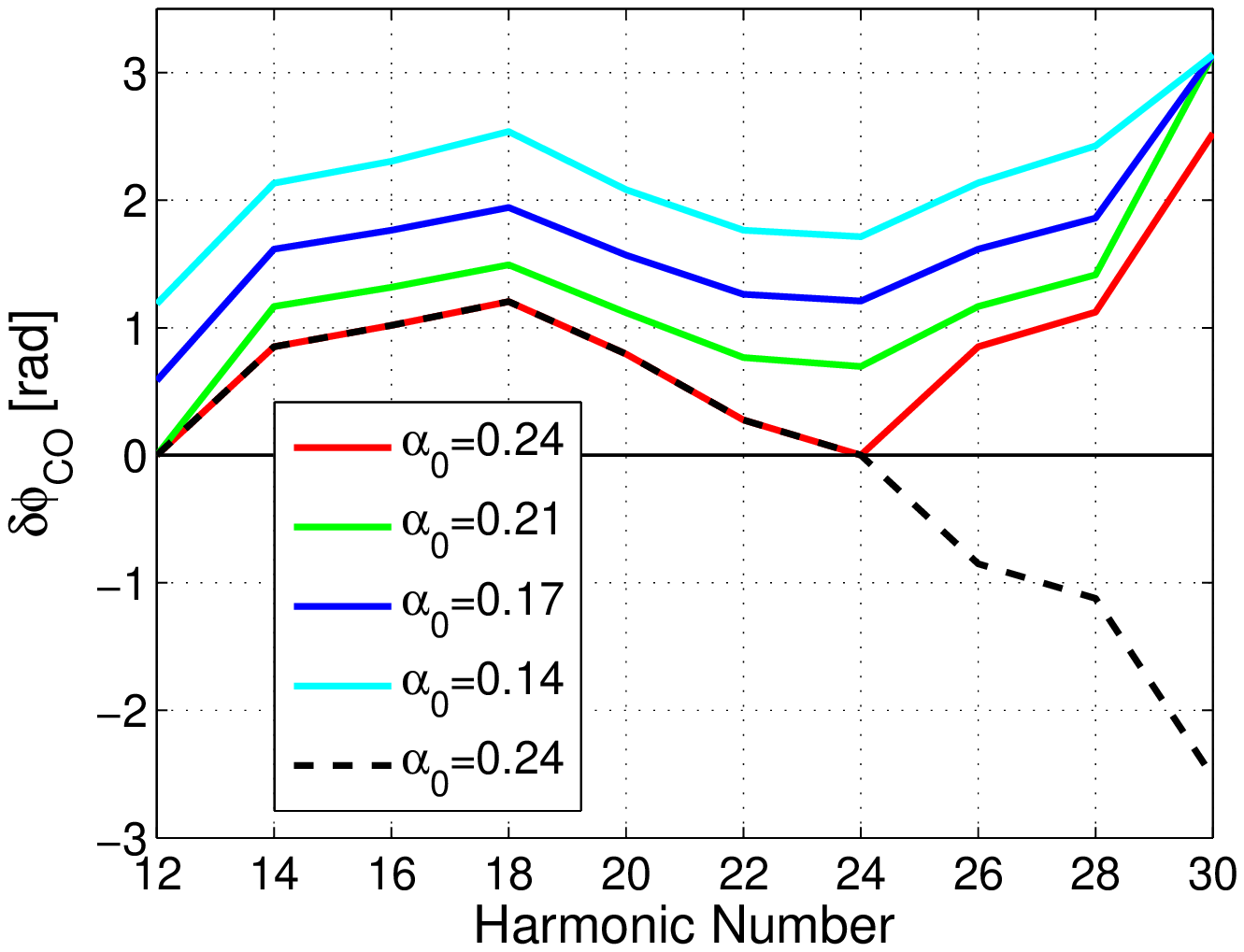}
	\caption{Phase reconstruction.}
	\label{phase_reconstruction}
\end{figure}

Since our estimated $\alpha_o$ has some uncertainty, we plot $\triangle\varphi(\nu_{2n})$ for a range of
$\alpha_o$ in Fig. \ref{phase_reconstruction}.
 In fact, the choice of possible values for the degree-of-orientation is restricted  within a range of $0.13-0.24$, because $\triangle\varphi(\nu_{2n})$ becomes imaginary outside of this range - $\alpha_o$ within this range is consistent with our initial estimates.  For all $\alpha_o$, the phase asymmetry increases to a maximum value near H18 ($\mathrm{h \nu \sim27 eV}$) and then decreases once again, exhibiting local minimum at H24. This local minimum is a robust feature, independent of the actual degree-of-orientation assumed.

\begin{figure}[t]
	\centering
	\includegraphics[width=\columnwidth]{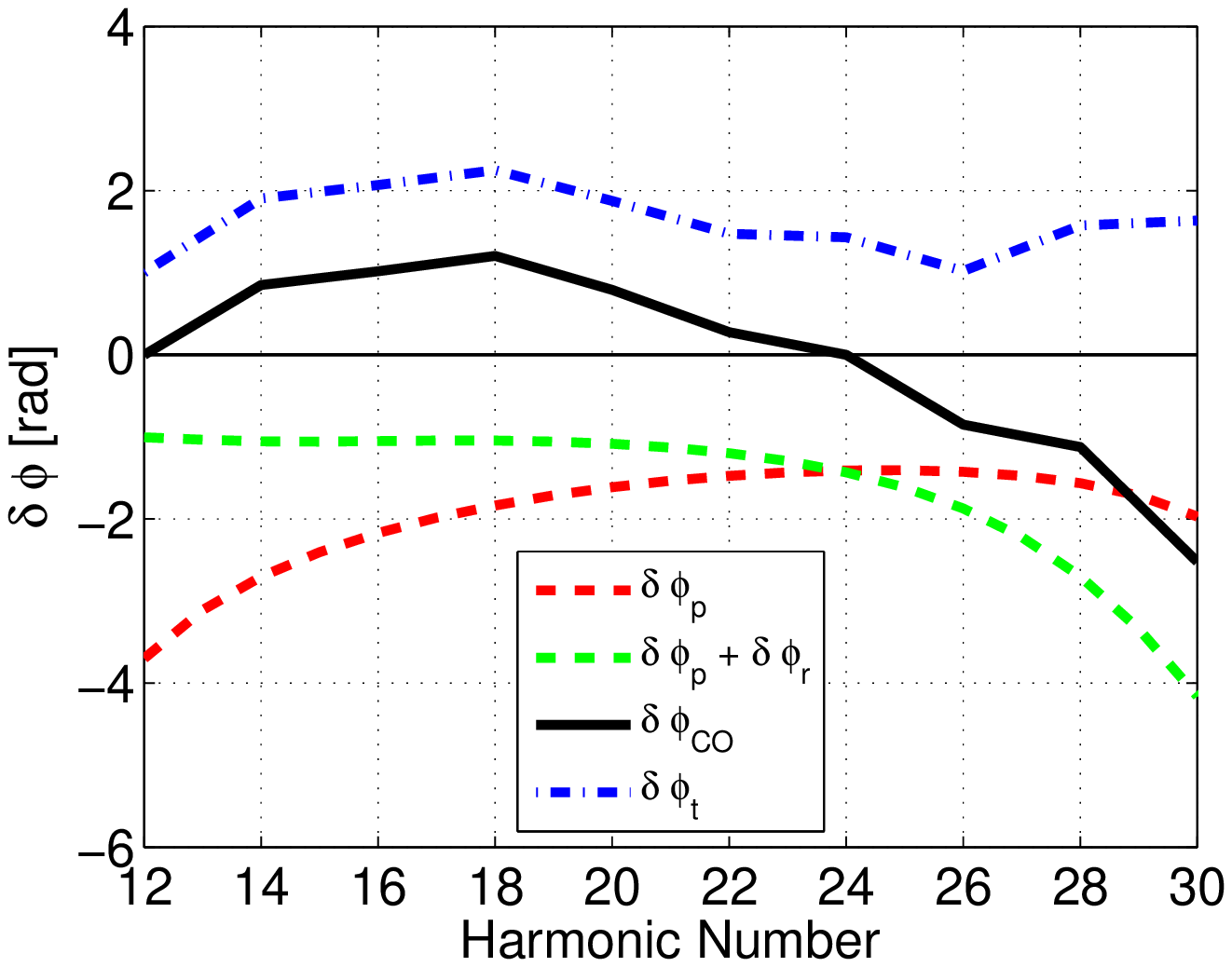}
	\caption{Phase asymmetry $\delta \phi= \phi_C-\phi_O$. The dashed red line depicts the propagation phase asymmetry, $\delta \phi_p$. The dashed green line shows the sum of propagation and recombination phases asymmetries $\delta \phi_p + \delta \phi_r$.
The solid black line shows the reconstructed phase asymmetry $\delta \phi_{CO}$. The dotted-dashed blue line depicts the estimated tunneling phase asymmetry $\delta \phi_t=\delta \phi_{CO} - \delta \phi_p - \delta \phi_r$.}
	\label{phase_asymmetry}
\end{figure}

  For $\alpha_o\simeq0.24$, the reconstructed phase touches the zero at H24. Both the positive (red line) and negative (dashed black line) values of the phase would produce the same interference signature in the even/odd harmonic ratio.
Local minima are common in many branches of physics and physical chemistry. They are often a signature of the monotonic phase changes.
Our estimates of the phase asymmetry acquired during propagation (using the strong field approximation) and recombination (dashed green line) show monotonic behavior with no local minimum (Fig. \ref{phase_asymmetry}).
This suggests that the total phase may be a monotonic function as well. For this reason we assume that the negative branch corresponds to physical reality.

In the laboratory frame the molecules are not perfectly aligned or oriented. We developed a more sophisticated model that takes into account averaging over alignment distributions. The results qualitatively are very similar to the simple model presented above - the overall shape of the curve, maximum (at H18) and minimum (at H24) positions are well reproduced.

Essentially the overall shape of the reconstructed phase (Fig. \ref{phase_asymmetry}), $\delta \phi_{CO}$ (black curve), is similar to the calculated phase asymmetry acquired during propagation and recombination, $\delta \phi_p + \delta \phi_r$ (dashed green curve), with both showing significant decrease towards the cut-off.
The difference between the two would correspond to a residual phase which we define as a tunneling phase, $\delta \phi_t=\delta \phi_{CO} - \delta \phi_p - \delta \phi_r$, as discussed in the body of the paper.

 \subsection{Delay in photo-ionization}

\begin{figure}[t]
	\centering
	\includegraphics[width=\columnwidth]{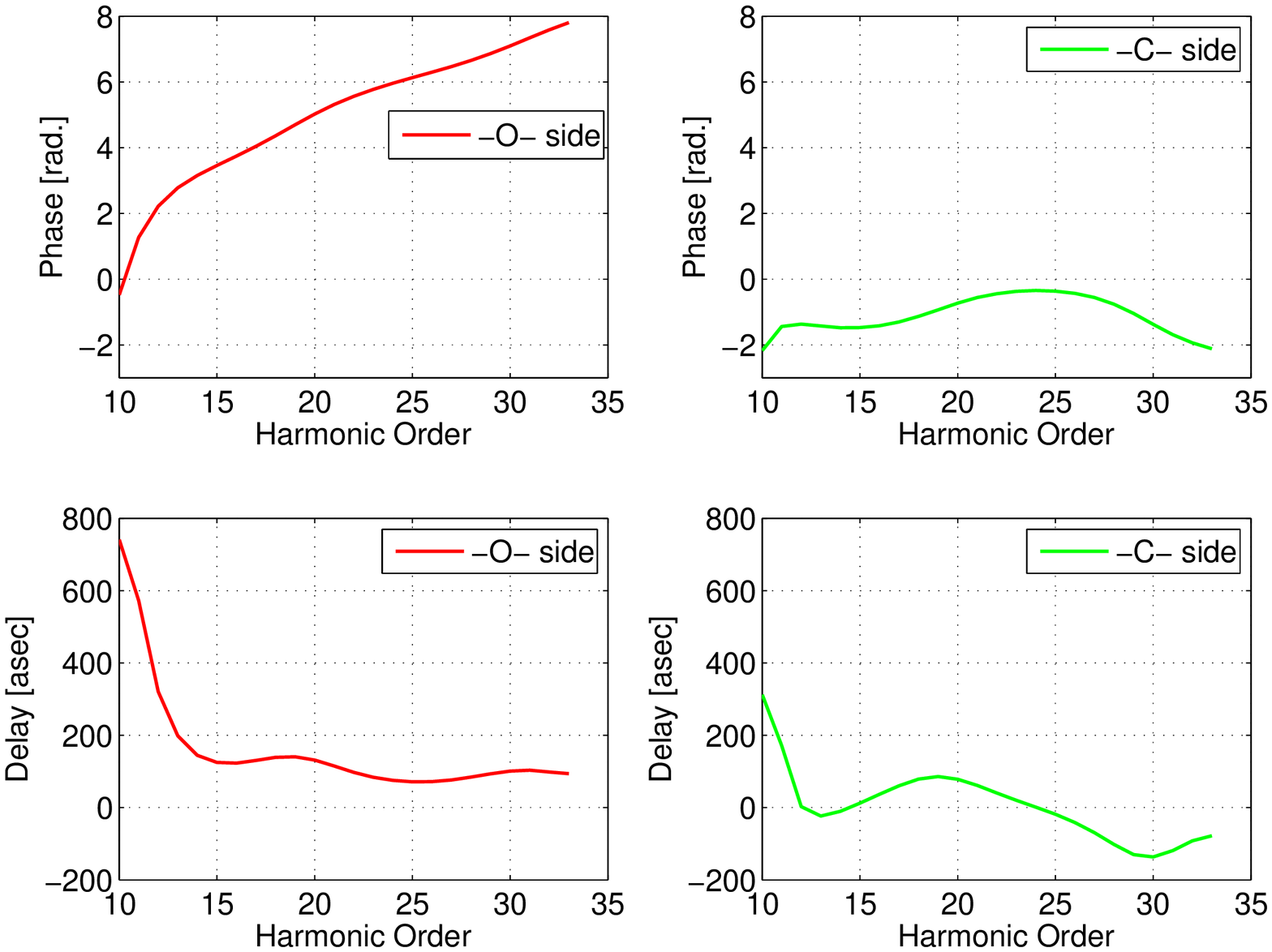}
	\caption{Two upper panels depict the phase of the transition moment along the intramolecular axis from the $-\mathrm{O}-$ and $-\mathrm{C}-$ sides. Two lower panels show time delay in photo-ionization from the $-\mathrm{O}-$ and $-\mathrm{C}-$ sides.}
	\label{time_delay}
\end{figure}

  The time delay ($\tau$) of photo-ionization relative to an attosecond pulse is proportional to the derivative of the phase with respect to energy ($\epsilon$) ($\tau=\hbar\partial\phi_r/\partial\epsilon$). This allows us to calculate the time delay from the phase structure of the transition moment along the intramolecular axis as a function of harmonic order from the $-\mathrm{O}-$ and $-\mathrm{C}-$ sides is shown on the two upper panels in Fig. \ref{time_delay} respectively.

